\documentclass[useAMS,usenatbib]{mn2e}
\usepackage{amssymb}
\usepackage{deluxetable}

\title[ALMA observations of 3C\,445 hot spot]{ALMA polarization observations of the particle accelerators in
  the hot spot of the radio galaxy 3C\,445}
\author[M. Orienti et al. ]
  {M. Orienti$^{1}$\thanks{E-mail: orienti@ira.inaf.it},
G. Brunetti$^{1}$, H. Nagai$^{2}$, R. Paladino$^{1}$, K.-H. Mack$^{1}$, M.A. Prieto$^{3}$\\
$^1$INAF -- Istituto di Radioastronomia, via Gobetti 101, I-40129 Bologna,
Italy \\
$^2$National Astronomical Observatory of Japan, Osawa 2-21-1,
Mitaka, Tokyo 181-8588, Japan\\
$^3$Instituto de Astrof\'isica de Canarias (IAC), E-38200 La Laguna, Tenerife, Spain}
\date{Received \today; accepted ?}

\pagerange{\pageref{firstpage}--\pageref{lastpage}} \pubyear{2002}

\def\LaTeX{L\kern-.36em\raise.3ex\hbox{a}\kern-.15em
    T\kern-.1667em\lower.7ex\hbox{E}\kern-.125emX}

\begin{document}

\label{firstpage}

\maketitle

\begin{abstract}
We present Atacama Large Millimeter Array (ALMA) polarization
observations at 97.5 GHz of the southern hot spot of the radio galaxy
3C\,445. The hot spot structure is dominated by two bright
components enshrouded by diffuse emission. Both components show
fractional polarization between 30 and 40 per cent, suggesting the presence
of shocks. The polarized
emission of the western component has a displacement of about 0.5 kpc
outward with respect to the total intensity emission, and may trace
the surface of a front shock. Strong polarization is observed in a
thin strip marking the ridge of the hot spot structure
visible from radio to optical. No significant
polarization is detected in the diffuse emission between
the main components,
suggesting a highly 
disordered magnetic field likely produced by turbulence and
instabilities in the downstream region that may be at the origin of
the extended optical emission observed in this hot spot. The
polarization properties support a scenario in
which a combination of both multiple and intermittent shock fronts due
to jet 
dithering, and spatially distributed stochastic second-order Fermi
acceleration processes
are present in the hot spot complex.
\end{abstract}


\section{Introduction}

\begin{figure}
\begin{center}
\includegraphics{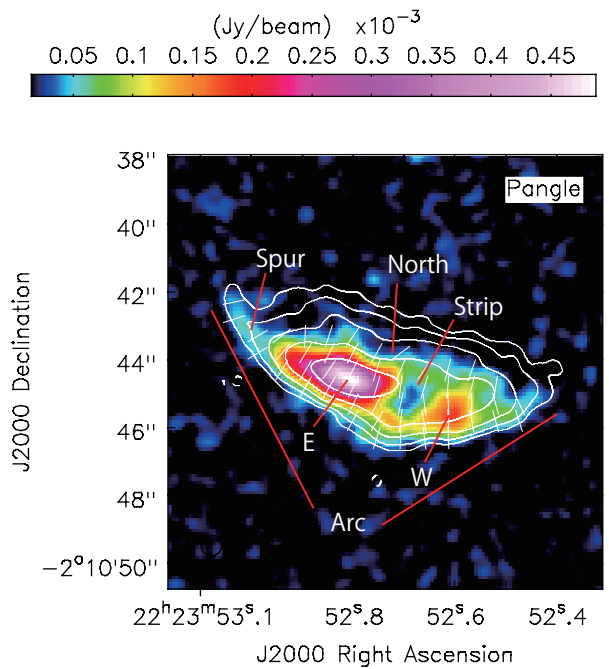}
\vspace{9.0cm}
\caption{ALMA image at 97.5 GHz of 3C\,445 South in total intensity
  (contours) overlaid on polarization intensity
  (colour-scale). The total intensity peak flux density is 1.3 mJy
  beam$^{-1}$. The first contour is 45 $\mu$Jy, which is 3 times the
  rms measured on the total intensity image.
Contours increase by a factor of 2. The polarization intensity colour
scale is shown by the 
  wedge at the top of the image. The 1-$\sigma$ noise level
  measured on the polarization image plane 
  is 12 $\mu$Jy beam$^{-1}$. Vectors
represent the electric vector position angle. The restoring beam is
plotted on the bottom left corner.}
\label{vector}
\end{center}
\end{figure}

Radio hot spots are bright and compact regions located at the
edges of powerful Fanaroff-Riley type II (FRII) radio galaxies
\citep{fanaroff74}. In the standard scenario hot spots are considered to be
the working surfaces of the supersonic jets launched by the active
galactic nucleus (AGN) which impact on the surrounding ambient
medium. In these regions a shock is produced, leading to the
acceleration of relativistic particles transported by the jet and
enhancing their emission. This scenario predicts very efficient
particle acceleration and has been strongly supported by the
detection of synchrotron optical/infrared emission from a number of
hot spots
\citep[e.g.][]{meisenheimer97,cheung05,mack09}. \\
In the standard shock model, particle acceleration occurs very
locally, implying that optical emission should arise from a very
compact region, up to a few hundred parsecs in size, due to the short
radiative lifetime of the energetic electrons
($\gamma > 10^{5}$).  
This simple shock scenario has been challenged by the discovery of
kpc-scale extended optical emission in a handful of hot spots, like 3C\,445
South \citep{prieto02}, 3C\,105 South \citep{mo12}, 3C\,445 North, 3C
227 West \citep{mack09}, Pictor\,A \citep{thomson95}, 3C\,33, 3C\,111,
3C\,303, 3C\,351 \citep{lahteenmaki99}, 3C\,390.3 \citep{prieto97},
3C\,275.1 \citep{cheung05}, and 4C\,74.26 \citep{erlund10}.\\
The discovery of kpc-scale diffuse optical emission suggests that
additional efficient and spatially distributed acceleration
mechanisms must take place in order to compensate the radiative losses
and to explain the extended synchrotron optical emission. The nature
of this additional acceleration process is still unclear. It may be
either multiple/complex shock surfaces in the hot spot region, or
efficient spatially distributed second-order acceleration processes
driven by
turbulence in the downstream region. The latter hypothesis
was proposed in the case of the hot spot 3C\,445 South \citep{prieto02}.\\
The peculiar hot spot 3C\,445 South is clearly detected from radio to X-rays
\citep{perlman10,mo12}. This hot spot has an east-west arc-shaped
structure of about 
10$\times$3 kpc in size with two main
compact components enshrouded by diffuse
emission visible in radio, NIR and optical bands.
In the optical another compact
component is present to the north-west of the brightest radio
component \citep{mo12}. Its location is marked as North in
Fig. \ref{vector}. The X-ray emission 
is misaligned with respect to the two main components and is more
consistent with the position of the compact optical
component \citep{perlman10,mo12}. 
\citet{prieto02} suggested that the development of
instabilities in the cocoon may produce
magneto-hydrodynamic turbulence that can account for the spatially
distributed reacceleration of the relativistic electrons that is
required to explain the diffuse optical emission between the compact
components. Such a second order acceleration mechanism becomes efficient
in presence of a low magnetic field (a few tens of milliGauss),
as the one estimated for 3C\,445 
South \citep{prieto02}. However, a clear evidence in
support of either the 
multiple/complex shock scenario or the turbulence model requires deep
polarimetric observations in the mm regime where contributions from
the steep-spectrum lobe and from the Faraday rotation are marginal. 
Alfv\'enic or super-Alfv\'enic turbulence is expected to tangle
efficiently the magnetic field on multiple scales and thus to reduce
significantly the fractional polarization. \\
In this Letter we present results on Atacama Large Millimeter Array
(ALMA) full-polarization observations at 97.5 GHz (3\,mm) of the hot
spot 3C\,445 
South. The polarization properties provide us with information on the
ordered component of the magnetic field and its topology
in the hot spot region. ALMA data are then complemented with Very
Large Array (VLA)
observations at 8.4 
GHz in order to trace the spectral index distribution. \\ 
This Letter is organized as follows: in Section 2 we describe the
observation and data reduction. In Section 3 we present results
on the ALMA data analysis, and in Section 4 we discuss the
polarization and spectral index properties in the context of particle
acceleration processes.\\
Throughout this paper, we assume the following cosmology: $H_{0} =
71\; {\rm km \; s^{-1} \; Mpc^{-1}}$, $\Omega_{\rm M} = 0.27$ and
$\Omega_{\rm \Lambda} = 0.73$, in a flat universe. 
At the redshift of the target, $z$ = 0.0558, 
the luminosity distance $D_{\rm L}$ is
252 Mpc, and 1 arcsecond = 1.09
kpc. \\

\section{Observations and Data Reduction}

\begin{figure}
\begin{center}
\includegraphics{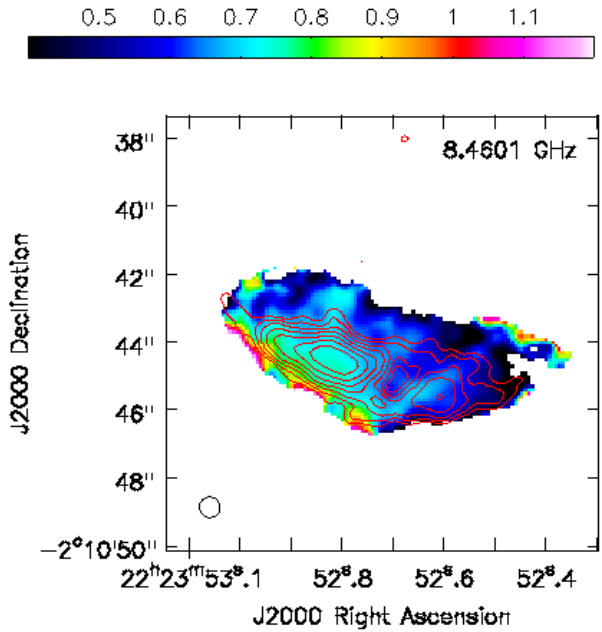}
\vspace{9.0cm}
\caption{Colour-scale spectral index image between 97.5 and 8.4 GHz of
  3C\,445 South superimposed on the ALMA polarization image at
  97.5 GHz (contours). The first contour is 45 $\mu$Jy 
  beam$^{-1}$, which corresponds to 3 times the rms measured on the
  image plane. Contour levels increase by a factor of $\sqrt{2}$. The
  colour scale is shown by the 
  wedge at the 
  top of the image. The restoring beam is plotted on
  the bottom left corner. }
\label{spix}
\end{center}
\end{figure}

We observed the hot spot 3C\,445 South with ALMA during Cycle 2 using
the Band 3 receiver (97.5 GHz) in full polarization (project ID:
\#2013.1.00286.S). The observations were performed on 2015
September 3 and 32 good antennas participated in the observing
run. The target on-source observing time was 2 hr, for a total
observing time of 3 hr that was required for ensuring a sufficient
parallactic angle coverage.
3C\,446 was observed as the bandpass and phase calibrator,
while Ceres was observed as the flux calibrator. However, it has been
found out that Ceres gives an inaccurate flux  
calibration, so the phase calibrator has been used instead, taking the flux
from the ALMA catalogue (1.322 Jy at 97.5 GHz). 
In addition, J2134-0153 was observed for calibration of the
instrumental polarization (D-terms), cross-hand delay, and cross-hand
phase.\\
Calibration and data reduction was done using Common Astronomical Software
Applications (CASA) version 4.7.0. Details of ALMA polarization data
calibration can be found in \citet{nagai16}. Final images have an rms of 15
$\mu$Jy/beam in Stokes I, and 12 $\mu$Jy/beam in Stokes Q and U. The
resolution is 0.6 arcsecond. No primary beam correction was
needed since the target was on-axis. Images in Stokes Q and U have
been used to produce the polarization intensity and polarization
angle images (Fig. \ref{vector}), as well as the fractional
polarization image and the associated statistical error image
(Fig. \ref{fractional}). 
Blanking on the fractional polarization image was
done by clipping the pixels of the total intensity image with values
below three times the rms measured on the off-source image
plane. \\
In addition to the ALMA data, we made use of VLA
observations at 8.4 GHz in A-configuration to perform the spectral
index analysis. Calibration and data
reduction of VLA data can be found in \citet{mo12}. We produced
another set of
VLA and ALMA
images with the same {\it uv}-range,
resolution and image sampling. Image registration was carried out by
comparing the peak of the brightest optically-thin hot spot
component. 
Then, we produced a spectral index
image\footnote{In this Letter we define the spectral index $\alpha$ as
$S_{\nu} \propto \nu^{- \alpha}$.} between 97.5 and 8.4 GHz. Blanking
was done by clipping the pixels
of the input images with values below three times the rms measured on
the off-source image plane at each frequency (Fig. \ref{spix}). Despite the particular
care used to produce and align the images, some gradients at the
edges are visible, due to the slightly larger extension of the
hot spot complex at the lower frequency.\\
The total intensity flux density at 8.4 and at 97.5 GHz, and the polarization
flux density at 97.5 GHz are extracted from the same region by selecting
polygonal areas. The values are reported in Table
\ref{tabella}. Spectral index errors are computed by assuming a 3 per cent
uncertainty on the amplitude scale at both frequencies, and
considering the error 
propagation theory. Errors on the polarization parameters were computed following 
\citet{fanti04}, and range from
about 0.5 per cent in the brightest 
regions up to 25$-$50 per cent in the ridge of the hot spot structure
with low signal-to-noise ratio (SNR) (see Section 3).\\ 

\section{Results}

\begin{figure*}
\begin{center}
\includegraphics{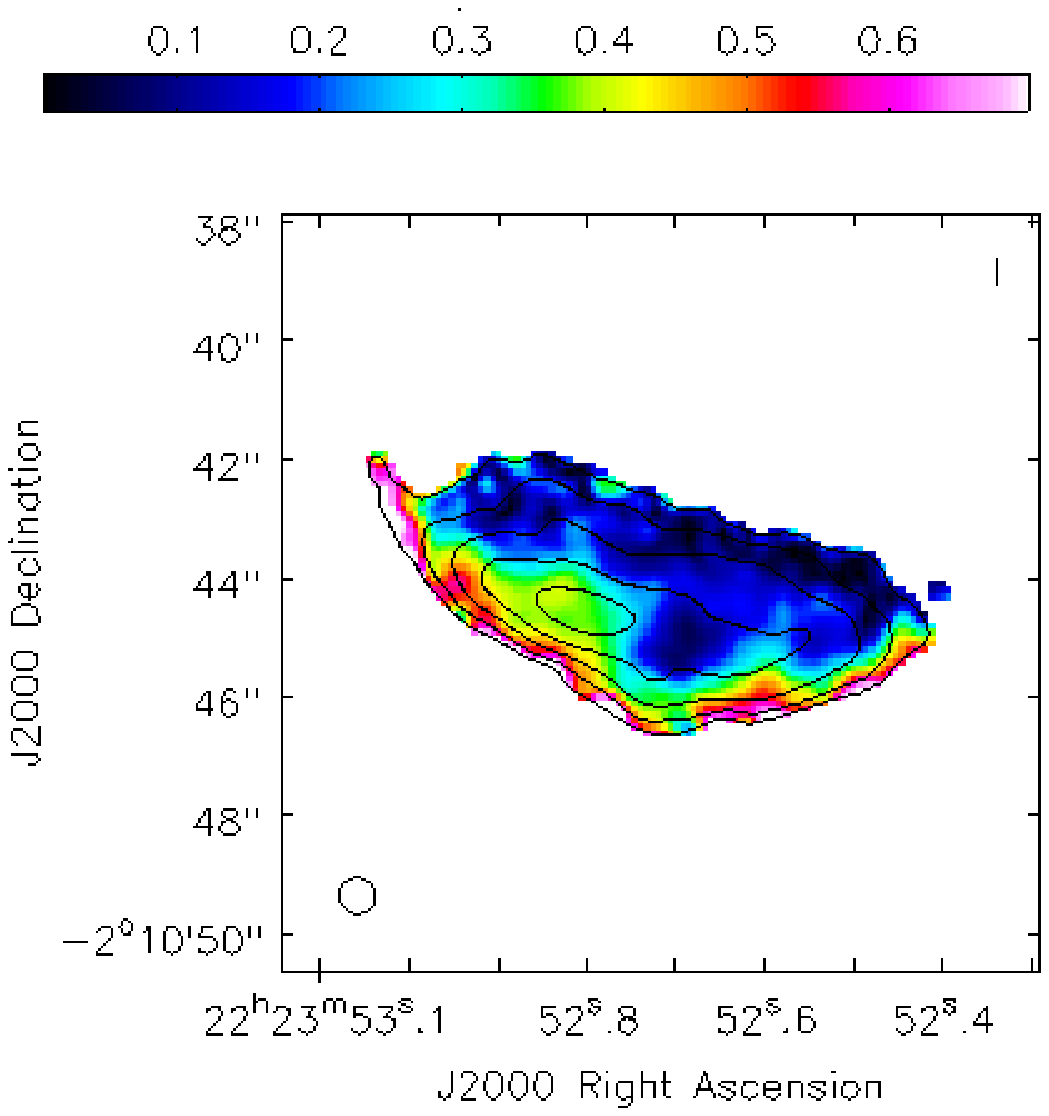}
\includegraphics{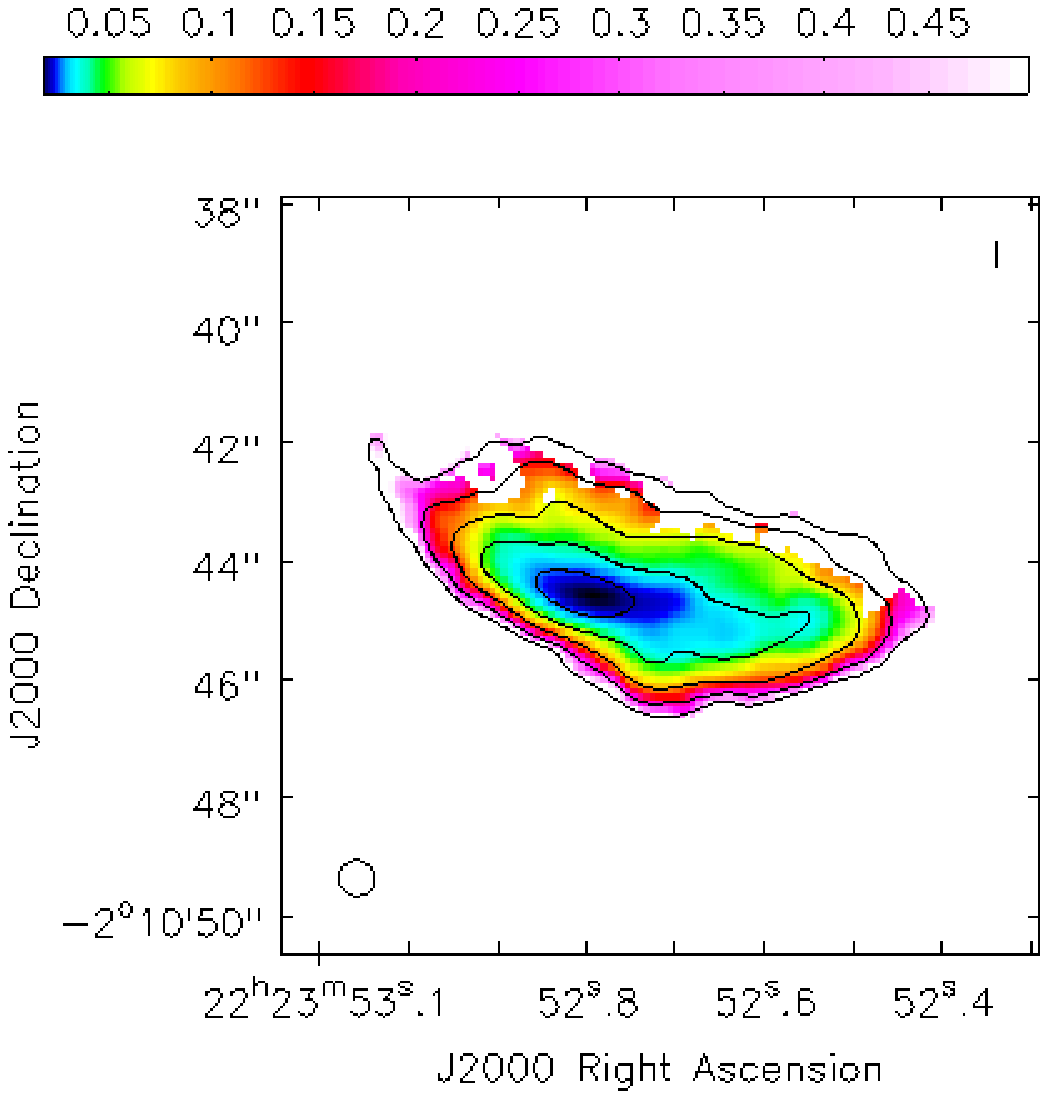}
\vspace{9.0cm}
\caption{ALMA image at 97.5 GHz of 3C\,445 South in total intensity
  (contours) overlaid on the fractional polarization ({\it left panel})
  and fractional polarization error ({\it right panel}) images
  (colour-scale). The total intensity peak flux density is 1.3 mJy 
  beam$^{-1}$. The first contour is 45 $\mu$Jy, which is 3 times the
  rms measured on the total intensity image.
Contours increase by a factor of 2. The fractional polarization and
the fractional polarization error colour
scales are shown by the 
  wedge at the top of the images. The restoring beam is
plotted on the bottom left corner.}
\label{fractional}
\end{center}
\end{figure*}

\begin{table*}
\caption{Observational parameters of the hot spot 3C\,445
  South. Column 1: hot spot component. Columns 2 and 3: total intensity flux density
  at 8.4 and 97.5 GHz, respectively. Column 4: polarization flux
  density. Column 5: polarization percentage. Column 6: spectral index between 97.5 and 8.4 GHz.}
\begin{center} 
\begin{tabular}{cccccc}
\hline
Comp. & S$_{8.4}$&S$_{97.5}$& $p$ & $m$ & $\alpha$\\
      & mJy     &  mJy   &  mJy & \% &  \\
\hline
E & 44.5$\pm$1.3 & 7.86$\pm$0.24 & 2.50$\pm$0.08& 32 &
0.70$\pm$0.02\\
W & 14.4$\pm$0.5 & 3.15$\pm$0.09 & 0.90$\pm$0.03& 29 &
0.62$\pm$0.02\\
Tot & 99.3$\pm$3.0 & 20.4$\pm$0.61& 5.61$\pm$0.17 & 27 & 0.64$\pm$0.02\\ 
\hline
\end{tabular}
\end{center}
\label{tabella}
\end{table*}

In the ALMA image at 97.5 GHz the hot spot 3C\,445 South shows an
elongated structure with two main components (labelled W and E in
Fig. \ref{vector}) enshrouded by
diffuse emission, in good
agreement with what was already found at cm-wavelengths and in
NIR/optical bands \citep{mo12}. The polarization percentage of the
whole hot spot is about 27 per cent. Although the percentage
integrated over the main components is 
about 30 per cent, it may reach peak values as high as 40 and
35 per cent in the E and W component, respectively
(Fig. \ref{vector}). Polarized emission is observed along the ridge of
the arc-shaped structure, with a polarization percentage from  
about 30 per cent up to 70 per cent in the north-eastern spur
(Fig. \ref{vector} and left panel of Fig. \ref{fractional}). The
uncertainty on the fractional polarization depends on the SNR and is
smaller for higher SNR (right panel of 
Fig. \ref{fractional}). 
The electric
vector position angles (EVPA) 
range between
-33$^{\circ}$ and -22$^{\circ}$ in the E component, and between
-5$^{\circ}$ and 15$^{\circ}$ in the W component
(Fig. \ref{vector}). There is a clear offset 
of about 0.5 arcsec (0.5 kpc) between the 
peak of the total intensity and the peak of polarized emission of the
W component, 
in the sense that the total intensity is nearer the nucleus.\\
A strip of polarized emission
of about 15 per cent is detected to the north of the two main
components (Figs. \ref{vector} and \ref{fractional}). The
compact component detected in optical/X-ray bands (labelled North in
Fig. \ref{vector}) is located in this
polarized region. 
No significant
polarization is observed from the diffuse emission ($<$3$\sigma$). In
particular, between E and W, we could set a 
tight upper limit on the fractional polarization of about 4 per cent,
which corresponds to the 
fractional polarization error computed in that region
(Fig. \ref{fractional}).\\
The spectral index between 97.5 and 8.4 GHz integrated over the whole
hot spot is $\alpha =$ 0.64$\pm$0.02 and is derived from the set of
images with the same {\it uv}-range, resolution and sampling at both
frequencies. The spectral index of the
E and W components
is $\alpha =0.70\pm 0.02$ and 0.62$\pm$0.02, respectively. The
spectral index of the diffuse emission ranges between 0.55 and 0.65. 
It is interesting to notice that between E and W, where
no significant polarization is detected, the spectral index is
0.60$\pm$0.02, and thus is slightly flatter than in the two compact and highly
polarized components.  \\

\section{Discussion and conclusions}

Synchrotron optical emission extending on kpc-scale is hard to reconcile
with the short radiative time of energetic optical-emitting electrons,
and indicates the 
presence of in-situ reacceleration processes. The
nature of these 
acceleration mechanisms, either Fermi-I processes like shocks,
or stochastic Fermi-II processes is still
unclear. \citet{prieto02} proposed a scenario in which
electrons are first
reaccelerated in a single region by strong shocks (Fermi-I), and then,
after being released from the shock region, 
are transported to the cocoon where they are further reaccelerated by
turbulence and compression. 
The
diffuse optical emission and the secondary W component were then
explained by 
means of stochastic Fermi-II mechanisms which become efficient in
presence of low 
magnetic fields ($\lesssim$ 100 $\mu$G) that prevent fast particle
cooling, like in the case of the 
low-luminosity hot spot of 
3C\,445 \citep{prieto02, brunetti03}.\\ 
ALMA observations in
the millimeter regime detect a high fraction of polarized emission from
various structures of the hot spot complex. In both E and W components
the fractional polarization reaches values of about 35 - 40 per cent. 
A polarization percentage between 30 and 70 per cent is
observed along the ridge of the 12-kpc arc-shaped structure, although
the low SNR prevents a tight constraint of its value.
Additional polarization of about 15 per cent is also 
observed in the northern strip bridging the E and W components
(Fig. \ref{fractional}). \\ 
These polarization percentages are too high to be reconciled with a
turbulent and uncompressed magnetic field. Conversely they
indicate the presence of 
multiple shock regions where the field is compressed in the direction
perpendicular to the shock fronts, one of them being coincident with component W. Following \citet{laing80} such high degrees of
  polarization may be observed when our line of sight is roughly parallel to
  the compression front, and thus forms a large
  angle to the jet axis ($\gtrsim 60^{\circ}$). In this case it is
  unlikely that the X-ray emission is produced by inverse Compton scattering of
  Cosmic Microwave Background photons in a decelerating jet
  \citep[e.g.][]{Georganopoulos04}. This supports the synchrotron
  radiation as the origin of the X-ray emission.
The EVPA are perpendicular to the highly polarized
structures, in agreement with what was found by \citet{leahy97} at 8.4
GHz, further supporting the presence of strong shocks. It is worth
mentioning that the peak of 
the polarized emission occurs about 0.5 kpc from the peak of the total
intensity, in a region where the spectrum is
slightly flatter ($\alpha \sim$0.63) than the spectral index measured
in the region of the 
total intensity peak ($\alpha \sim$0.69). This result is in agreement
with the idea that the polarization 
marks the shock front, while the total intensity emission traces also
backflow material.\\
The large-scale shock surface traced by the polarization marks the
largest scales where the kinetic energy of the jet is dissipated.  
The multiple shock surfaces (i.e. the E and W components imaged from radio
to optical band, as well as the North component detected in the
optical) suggest either that the jet may change 
its direction, creating a dentist drill effect, or that the shock
surface is very complicated. 
Changes in the jet
directions may be caused by instabilities as the jet
interacts with backflow material
\citep[e.g.][]{perucho14,tchekhovskoy16}.
In principle, shocks would accelerate electrons to very high energies
and this has been used to highlight some tension between shock
acceleration models and the optical synchrotron cut-off observed in
some hot spots \citep{araudo16}. Given our results this tension
may be circumvent by assuming that the shock acceleration occurs in an
intermittent way. It is possible that the E and W components mark
regions where particle acceleration occurred 100-1000 yr ago, while
the North component that emerges in the optical band and that
is probably emitting synchrotron radiation up to X-rays, pinpoints
acceleration that is taking place at the current epoch. A similar
scenario has been proposed for explaining the X-ray emission in the
hot spot of Pictor\,A. \citet{tingay08} suggested that X-rays in
Pictor\,A are produced by synchrotron emission from compact pc-scale
transient regions of enhanced magnetic field. This is consistent also
with the recent discovery of temporal variability of the 
X-ray emission in the hot spot of Pictor\,A \citep{hardcastle16}. \\
An important result obtained by our ALMA observations of the hot spot
3C\,445 South is the lack of significant polarization between E
and W and in the diffuse emission located to the north of
the polarized emission and 
visible up to the optical I-, R-, and B-band \citep{mo12}.
Due to the lack of polarization the magnetic field in this region
should be tangled on scales that are much smaller than the beam
equivalent scale, i.e. 600 pc. Specifically, assuming a simple picture
where the turbulent magnetic field produces random polarization
orientation, the size of the turbulent cells should be 

\begin{equation}
l_{\rm t} \sim \left( \frac{p_{\rm obs}}{p_{\rm int}}
    \right)^{2/3} \left( \frac{A_{\rm b}}{\rm 0.3 kpc^2}
      \frac{\phi}{\rm 4
      kpc} \right)^{1/3} {\rm kpc}
\label{equation}
\end{equation}

\noindent where $p_{\rm obs}$ and $p_{\rm int}$ are the observed and the
intrinsic polarization percentage, respectively, $A_{\rm b}$ is the
beam area, and $\phi$ is the size of the turbulent region. If in
Eq. \ref{equation} we consider $p_{\rm int} \sim 70$ per cent and $p_{\rm obs} < 4$
per cent, we obtain $l_{\rm t} \sim$ 160 pc. 
Particles injected in E and W
are then transported and possibly reaccelerated downstream in
a turbulent magnetic field by Fermi-II mechanisms\footnote{The scale
  derived from polarization is
  in line with the turbulence scales, $\sim$ 20 pc, that were derived
  by \citet{prieto02} to explain optical emission via Fermi-II acceleration.}.
The spectrum of the diffuse emission, which is
flatter than in the main shock fronts, may support this interpretation.\\
In the hot spot of 3C\,445 the detection of strong shocks occurring in
different regions, as well as efficient stochastic Fermi-II processes taking
place in the diffuse emission clearly indicates a complex distribution and
nature of particle acceleration at the edge of radio
galaxies. Furthermore, it provides a significant 
modification to the classical scenario of shock acceleration that is routinely adopted for hot spots. \\
Further high-frequency full-polarization observations of hot spots with
diffuse optical emission will be useful for determining the incidence
of the various acceleration mechanisms and their distribution across
the hot spot structure. \\

\section*{Acknowledgments}
This paper makes use of the following ALMA data:
ADS/JAO.ALMA\#2013.1.00286.S. ALMA is a partnership of ESO
(representing its member states), NSF (USA) and NINS (Japan), together
with NRC (Canada), NSC and ASIAA (Taiwan), and KASI 
(Republic of Korea), in cooperation with the Republic of Chile.
The Joint ALMA Observatory is operated by ESO, AUI/NRAO and NAOJ.
The VLA is operated by the US National Radio Astronomy Observatory
which is a facility of the National Science Foundation operated under
cooperative agreement by Associated Universities, Inc. 
Part of this work was done with the contribution of the Italian
Ministry of Foreign Affairs and Research for the collaboration project
between Italy and Japan. HN is supported by MEXT KAKENHI Grant Number
15K17619.

\end{document}